# High mobility and high on/off ratio field-effect transistors based on chemical vapor deposited single-crystal $MoS_2$ grains

Wei Wu[1,2], Debtanu De[3,4], Su-Chi Chang[2,5], Yanan Wang[1], Haibing Peng[3,4], Jiming Bao[1] and Shin-Shem Pei[1,2,a]

[1] Department of Electrical and Computer Engineering, University of Houston, Houston, TX 77204, USA

[2] Center for Advanced Materials, University of Houston, Houston, TX 77204, USA

[3] Department of Physics, University of Houston, Houston, TX 77204, USA

[4] Texas Center for Superconductivity, University of Houston, Houston, TX 77204, USA

[5] Materials Engineering Program, University of Houston, Houston, TX 77204, USA

We report field-effect transistors (FETs) with single-crystal molybdenum disulfide ($MoS_2$) channels synthesized by chemical vapor deposition (CVD). For a bilayer $MoS_2$ FET, the mobility is ~17 $cm^2V^{-1}s^{-1}$ and the on/off current ratio is ~$10^8$, which are much higher than those of FETs based on CVD polycrystalline $MoS_2$ films. By avoiding the detrimental effects of the grain boundaries and the contamination introduced by the transfer process, the quality of the CVD $MoS_2$ atomic layers deposited directly on $SiO_2$ is comparable to the best exfoliated $MoS_2$ flakes. It shows that CVD is a viable method to synthesize high quality $MoS_2$ atomic layers.

---

[a] Author to whom correspondence should be addressed. Electronic mail: spei@uh.edu

The single-layer (SL) graphene has a linear Dirac-like band structure with no bandgap, which leads to the formation of massless Dirac fermions with remarkable electronic properties, e.g., an effective speed of light $v_F \approx 10^6$ ms$^{-1}$ and a room temperature mobility of 200,000 cm$^2$V$^{-1}$s$^{-1}$. However, the lack of a bandgap also limits the application of graphene. Recently, transition metal dichalcogenides (TMDs), molybdenum disulfide ($MoS_2$) in particular, have attracted a lot of attention. The bulk $MoS_2$ is a semiconductor with an indirect bandgap of ~1.3 eV and the SL $MoS_2$ has a direct bandgap ~ 1.8 eV.[1,2,3] Therefore, $MoS_2$ could complement graphene for many electronic and photonic applications. However, studies of mechanically exfoliated $MoS_2$ on $SiO_2$ found the room temperature mobility is < 10 cm$^2$V$^{-1}$s$^{-1}$ for SL-$MoS_2$ and 10~15 cm$^2$V$^{-1}$s$^{-1}$ for bilayer $MoS_2$,[4,5] which are substantially lower than the measured ~ 200 cm$^2$V$^{-1}$s$^{-1}$ of the bulk $MoS_2$,[6] or the calculated ~ 410 cm$^2$V$^{-1}$s$^{-1}$ of intrinsic n-type SL-$MoS_2$, which is limited only by optical phonon scattering.[7] The lower than expected mobility is partially due to the long ranged charge disorder or short ranged disorder caused by chemical bonding or roughness at the interfaces.[8] Furthermore, the mechanical exfoliation process cannot be scaled up for practical applications.

Only recently, large-area of SL and few-layer $MoS_2$ films have been synthesized by chemical vapor deposition (CVD),[9,10] sulfurization of $MoO_3$,[11] or thermolysis of $(NH_4)MoS_4$.[12] CVD has been demonstrated as the most practical method of synthesizing large-area and high quality graphene,[13] boron nitride[14] and BCN nanosheets.[15] However, devices fabricated from these polycrystalline $MoS_2$ films are still substantially inferior to their exfoliated counterparts.[4,16] One possible cause of the degradation of performance is the detrimental effects of the grain boundaries, which can be avoided in the case of graphene by going to a seeded growth single-crystal array approach by CVD to place graphene grains at predetermined locations where devices will be located.[17]

In this paper, we report the construction of field-effect transistors (FETs) based on single-crystal bilayer and few-layer $MoS_2$ grains. SL, bilayer and few-layer grains with sizes up to 20 μm were synthesized directly on $SiO_2$ by CVD. Bilayer and few-layer FETs offer higher on-state current than the SL-$MoS_2$ FET, while maintain high on/off current ratios.[5] With a single-crystal bilayer $MoS_2$ conducting channel, we have achieved a superior mobility of 17.3 $cm^2V^{-1}s^{-1}$ and a current on/off ratio of $4\times10^8$ in a back-gated $MoS_2$ FET.

Our CVD-growth method of single-crystal $MoS_2$ grains is a modification of what is described in Ref. [10] for continuous $MoS_2$ films. However, we do not use seeds as nucleation centers to initiate the growth. Single-crystal $MoS_2$ grains were synthesized in a conventional horizontal quartz tube furnace with sulfur and $MoO_3$ powders as source materials. The $MoO_3$ (0.1 g, Alfa, 99.5%) was placed in an alumina boat and loaded into the center uniform-temperature zone of the furnace. However, we found the residues deposited on the wall of the quartz tube furnace also contribute to the subsequent $MoS_2$ growth, which is not the focus of this paper and will be discussed in detail in another paper.

A piece of Si wafer with 300 nm $SiO_2$ layer was put downstream in a separate boat as substrate. Another alumina boat with 0.4 g sulfur (Alfa, 99.5%) was placed upstream in a low-temperature zone. Before growth, the furnace was evacuated down to ~70 mTorr and back-filled with Ar gas to ambient pressure. In the flow atmosphere of 100 sccm Ar, the furnace was heated to 700 °C at the center zone in 60 min subsequently up to 1100 °C in 130 min. The temperature of the sulfur and the substrate was increased concurrently to ~100 °C and ~700 °C, respectively. After 20 min, the furnace was cooled down naturally to room temperature.

Raman spectroscopy is used as a non-destructive method to characterize crystalline quality and thickness of MoS$_2$ grains. Representative Raman spectra of SL and bilayer MoS$_2$ grains are shown in Fig. 1. For MoS$_2$ crystals, two characteristic Raman active modes, $E_{2g}^1$ and $A_{1g}$, are found. They are associated with the in-plane and out-of-plane vibration of sulfides, respectively.[18] It has been reported that the peak frequency difference between $E_{2g}^1$ and $A_{1g}$ ($\Delta$) can be used to identify the number of MoS$_2$ layers.[9,11,19] Figures 2(a) and 2(b) show Raman intensity mappings of $E_{2g}^1$ at 383 cm$^{-1}$ and $A_{1g}$ at 405 cm$^{-1}$ of a triangular shape MoS$_2$ grain, which confirms the thickness and quality uniformity of the CVD grains. A $\Delta$ of 22 cm$^{-1}$ suggests the grain is a bilayer MoS$_2$ crystal. For SL MoS$_2$, $\Delta$ = 18 cm$^{-1}$ in our system. In Fig. 3, a typical photoluminescence (PL) spectrum of the bilayer grain presents two emission peaks at 676 nm and 630 nm, known as A1 and B1 direct excitonic transitions, respectively.[20] The PL result is also consistent with recent studies of large-area CVD MoS$_2$ films.[10,11]

The Individual MoS$_2$ grains were first visually inspected and selected under an optical microscope and their positions were recorded with respected to predefined marks. The numbers of MoS$_2$ layers of individual grains were determined by Raman spectroscopy. Subsequently, MoS$_2$ grains were fabricated into back-gated FETs with the standard microelectronics processes following steps similar to those described in Ref. 21. The patterned drain and source metal contact electrodes of 45 nm Pd (on top of a 5 nm adhesion layer of Cr) were fabricated on the selected MoS$_2$ grains by electron-beam lithography and a lift-off process.

Figure 4(a) shows an optical microscopy image of the FET under study in Figs. 4(b)-(d). The channel of the FET is bilayer MoS$_2$ determined by Raman spectroscopy. The degenerately doped Si substrate, which is separated from the $MoS_2$ channel by a 300 nm $SiO_2$, is used as a back gate to tune the charge carrier density in the $MoS_2$ channel via the application of a back gate voltage $V_G$. Room temperature electrical measurements were performed under vacuum ($10^{-5}$–$10^{-6}$ Torr) in a Lakeshore TTP6 cryogenic probe station.

Figure 4(b) shows the drain current $I_{DS}$ at fixed drain–source voltage, $V_{DS}$=+500mV, as a function of the applied back-gate voltage $V_G$, for the device shown in Fig. 4(a). The device is an n-channel normally-on FET. The field-effect mobility is determined using the formula: $\mu = (L/WC_{ox})\Delta G/\Delta V_G$,[22] where $G = I_{DS}/V_{DS}$ is the conductance and $\Delta G/\Delta V_G = (1/V_{DS})(\Delta I_{DS}/\Delta V_G)$ is determined from the slope of a linear-fit of the data with the back-gate voltage ranges from $V_G$=+80V to $V_G$=+100V. L = 1 μm is the length and W = 3.6 μm is the width of the $MoS_2$ channel determined from Fig. 2(a). $C_{ox} = \varepsilon_0\varepsilon_r/d$ is the capacitance per unit area, where d = 300 nm is the thickness of the $SiO_2$ layer with $\varepsilon_0$ = 8.854x$10^{-12}$ F$m^{-1}$ being the free-space permittivity and $\varepsilon_r$ = 3.9 being the relative permittivity of $SiO_2$. The field-effect mobility of the CVD bilayer $MoS_2$ is determined to be 17.3 $cm^2V^{-1}s^{-1}$ comparing to the previously reported 0.02 $cm^2V^{-1}s^{-1}$ of CVD SL-$MoS_2$,[10] and 0.04 $cm^2V^{-1}s^{-1}$ of the CVD few-layer $MoS_2$.[9] The much higher mobility of our device may be partially due to the elimination of grain boundary scattering as we reported previously for the CVD graphene.[17] Actually, the 17.3 $cm^2V^{-1}s^{-1}$ mobility of the CVD bilayer $MoS_2$ grain is comparable to the 0.1-10 $cm^2V^{-1}s^{-1}$ reported for exfoliated SL-$MoS_2$,[4] and the 10-15 $cm^2V^{-1}s^{-1}$ for exfoliated bilayer $MoS_2$.[5] Another order of magnitude improvement is expected if a high-κ dielectric is applied on the top of the $MoS_2$ channel to reduce the Coulomb effect.[4,5,23]

In Fig. 4(c), the drain current $I_{DS}$ is re-plotted on a logarithmic scale as a function of $V_G$. At $V_G$ = -100 V, the $MoS_2$ channel of the FET is pinched off with an off-state $I_{DS}$ < 0.1 pA. The on-state $I_{DS}$ is > 30 μA with $V_G$ = +100 V. The corresponding on/off current ratio is $4\times10^8$, which is higher than the ~ $10^4$ on/off current ratio reported for CVD polycrystalline $MoS_2$ films,[10] and comparable to the ~ $10^8$ of the exfoliated SL- $MoS_2$ flakes.[4]

Figure 4(d) shows the room temperature transfer characteristics of the FET, i.e. the dependence of drain current on the back-gate voltage at various drain-source voltages. Due to the thick $SiO_2$ back-gate dielectric, no drain current saturation is observed. For comparison, another back-gated FET with a few-layer (<5 layers) $MoS_2$ channel was also fabricated, the mobility is also ~ 17 $cm^2V^{-1}s^{-1}$, while the on/off current ratio might be slightly lower, but still > $10^4$. Most recently, ven der Zande et al. also reported the electrical characteristics of CVD single-crystal $MoS_2$ grains. The mobility measured within a grain was reported to be 3-4 $cm^2V^{-1}s^{-1}$ and the on/off current ratio was in the range of $10^5$ to $10^7$.[24] Our results are consistent with their findings.

It is well known that the best reported mobility of graphene on $SiO_2$ is limited to 10,000 $cm^2V^{-1}s^{-1}$ primarily due to the Coulomb effect.[25] For exfoliated multilayer $MoS_2$ on $SiO_2$, the room temperature mobility can be substantially enhanced by engineering the dielectric environment. For example, multilayer $MoS_2$ has exhibited a mobility > 100 $cm^2V^{-1}s^{-1}$ when sits on a 50-nm thick atomic layer deposited (ALD) $Al_2O_3$,[16] and 470 $cm^2V^{-1}s^{-1}$ on 50-nm thick spin-coated PMMA.[8] Further enhancement of the $MoS_2$ mobility can be achieved by applying appropriate gate dielectric on the top of $MoS_2$ channel. A mobility as high as ~200 $cm^2V^{-1}s^{-1}$ was achieved with a $HfO_2$/SL-$MoS_2$/$SiO_2$ structure, which also exhibits a high on/off ratio (~$10^8$) and low subthreshold swing (~70 mV per decade).[4] Thus, in addition to fundamental scientific interests, $MoS_2$ FETs could be an attractive candidate for low power electronics, e.g. thin-film transistors (TFTs) in the next generation high-resolution liquid crystal (LCD) or organic light-emitting diode (OLED) displays.[26]

In conclusion, we report the electrical characteristics of back gated FETs fabricated on single-crystal $MoS_2$ grains synthesized by CVD on $SiO_2$. A FET with a bilayer $MoS_2$ channel has a mobility ~ 17 $cm^2V^{-1}s^{-1}$ and an on/off current ratio ~ $10^8$, while the FET with a few-layer $MoS_2$ channel exhibits comparable mobility, but slightly lower on/off current ratio. Another order of magnitude improvement of mobility is expected by dielectric engineering to reduce the Coulomb effect. The results suggest that CVD is a viable method to synthesize high quality $MoS_2$ grains with performance comparable to the best mechanically exfoliated $MoS_2$ flakes, and $MoS_2$ FETs are promising candidates for low power electronics.

**FIGURES**

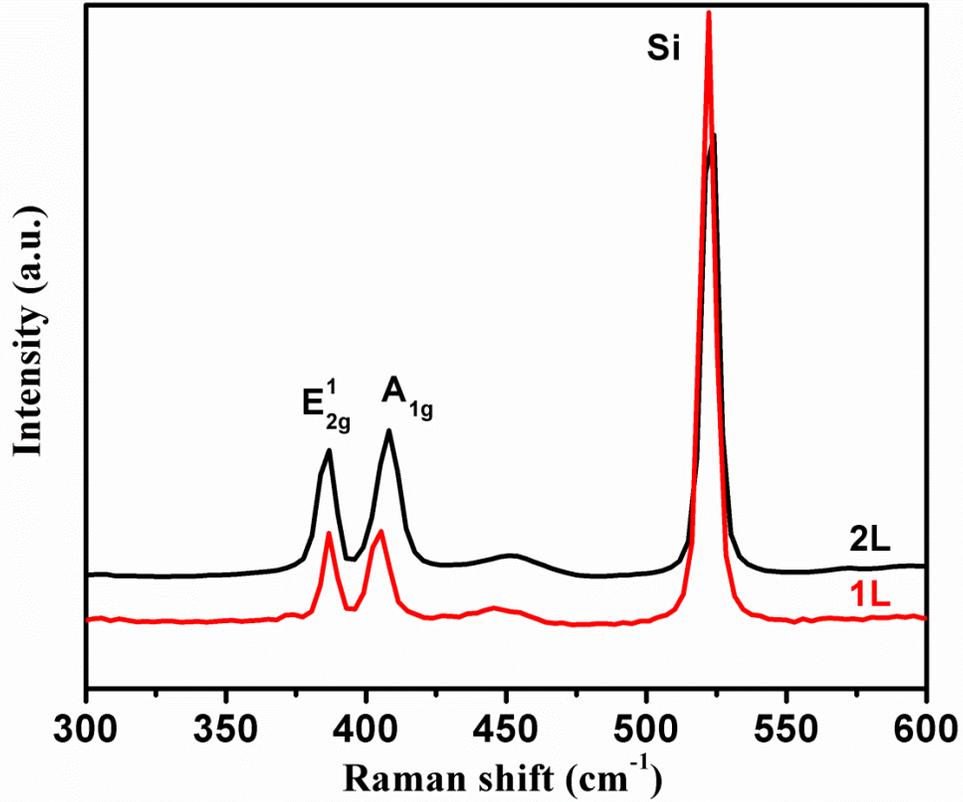

FIG 1. Raman spectra of typical single-layer and bilayer $MoS_2$ crystals. $E_{2g}^1$ at 383 cm$^{-1}$ and $A_{1g}$ at 405 cm$^{-1}$ for bilayer; $E_{2g}^1$ at 384 cm$^{-1}$ and $A_{1g}$ at 402 cm$^{-1}$ for single layer. The laser excitation wavelength is 532 nm.

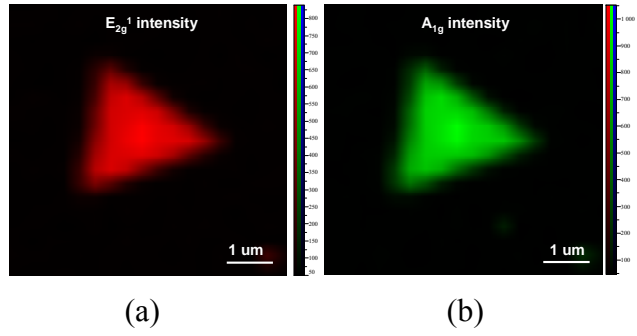

FIG. 2. Raman intensity mappings of (a) $E_{2g}^1$ and (b) $A_{1g}$ of a typical bilayer $MoS_2$ grain.

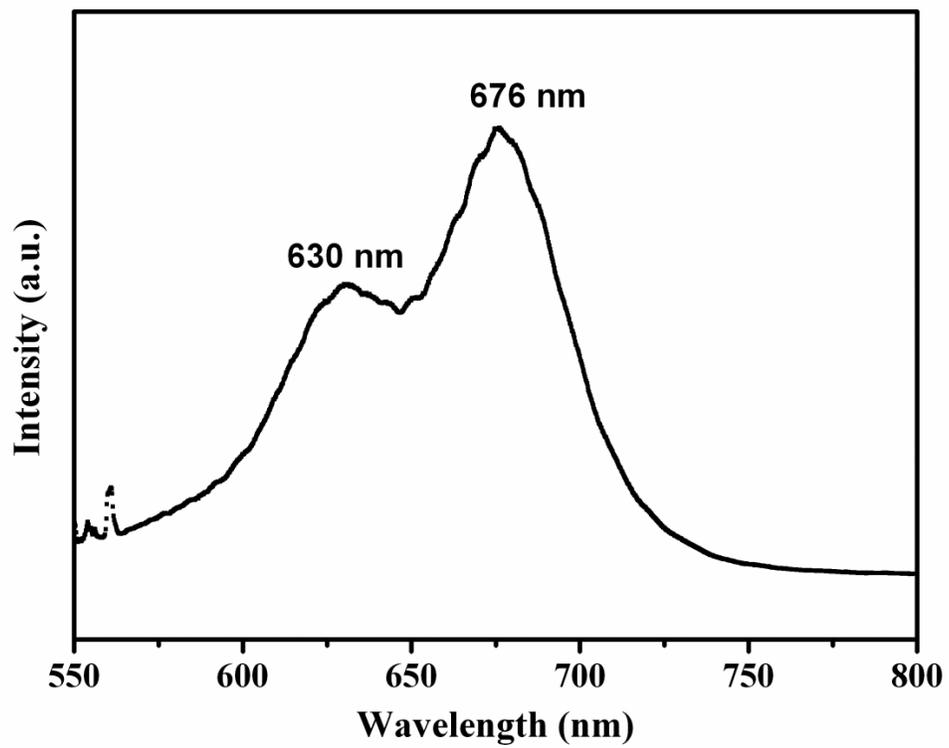

FIG. 3. Photoluminescence spectrum of a typical bilayer $MoS_2$ crystal. The laser excitation wavelength is 532 nm.

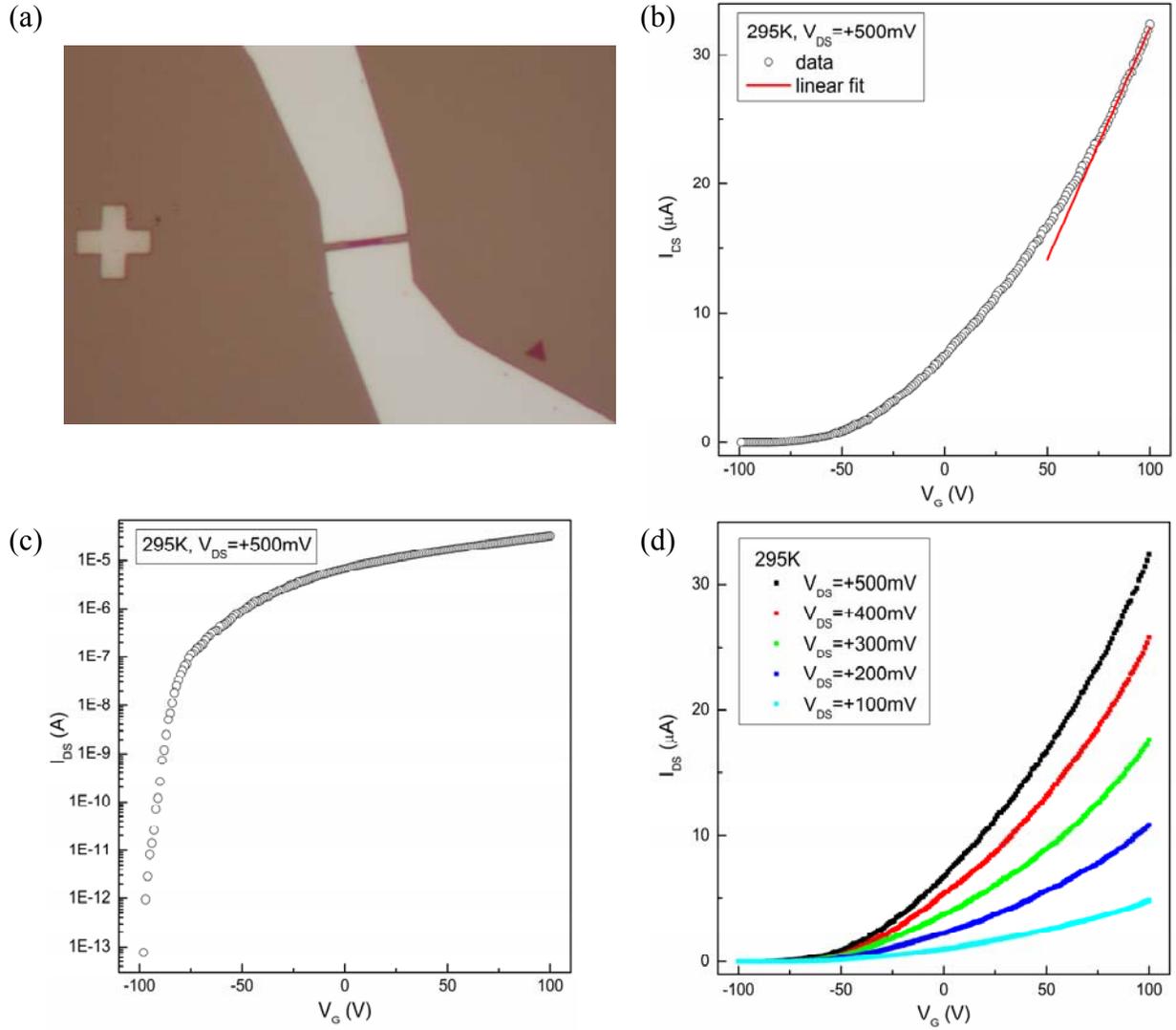

FIG. 4. (a) Optical image of the device. The gap between the two electrodes acrossing the MoS$_2$ grain is 1μm. (b) (Open circles) Drain-source current I$_{DS}$ as a function of back-gate voltage V$_G$ at fixed drain-source bias voltage V$_{DS}$=+500mV. (Red line) Linear-fit of the data within the back-gate voltage range from V$_G$=+80V to V$_G$=+100V. From the linear fit data, the carrier mobility is calculated to be μ=17.3 cm$^2$V$^{-1}$s$^{-1}$. (c) Drain-source current I$_{DS}$ plotted in logarithmic scale as a function of back-gate voltage V$_G$ at fixed drain-source bias voltage V$_{DS}$=+500mV. The optimized current pre-amplifier gain used in the measurement: 100pA/V for V$_G$=-100V to -90V, 10nA/V for V$_G$=-89V to -80V, 500nA/V for V$_G$=-79V to -40V and 10μA/V for V$_G$=-40V to +100V. (d) Drain-source current I$_{DS}$ as a function of back-gate voltage V$_G$ at drain-source bias voltages V$_{DS}$=+500mV, =+400mV, =+300mV, =+200mV and =+100mV.